\documentclass[12pt,a4paper,final]{iopart}

\usepackage{graphicx}
\usepackage[breaklinks=true,colorlinks=true,linkcolor=blue,urlcolor=blue,citecolor=blue]{hyperref}
\usepackage{amsfonts, amsmath, dsfont}
\usepackage{cite}

\def\epp{\: .}
\def\epc{\: ,}
\newcommand{\sine}[2]{s_{#1,#2}}

\begin{document}

\title[Overlap of the N{\'e}el state with XXZ Bethe states]{A Gaudin-like determinant for overlaps of N{\'e}el and XXZ Bethe states}

\author{M. Brockmann$^{1}$, J. De Nardis$^{1}$, B. Wouters$^{1}$, and J.-S. Caux$^{1}$}
\address{$^1$Institute for Theoretical Physics, University of Amsterdam, Science Park 904,\\
Postbus 94485, 1090 GL Amsterdam, The Netherlands}
\ead{M.Brockmann@uva.nl}

\begin{abstract}
We derive a determinant expression for overlaps of Bethe states of the XXZ spin chain with the N{\'e}el state, the ground state of the system in the antiferromagnetic Ising limit. Our formula, of determinant form, is valid for generic system size. Interestingly, it is remarkably similar to the well-known Gaudin formula for the norm of Bethe states, and to another recently-derived overlap formula appearing in the Lieb-Liniger model.
\end{abstract}


\section{Introduction}

In the last few decades one-dimensional quantum integrable models have proved to be immensely fertile theoretical laboratories for studying nonperturbative effects in strongly-correlated systems. It however remains very arduous to obtain results going beyond formal expressions for eigenfunctions and basic spectral properties. Significant further progress in the field has been obtained based on the fundamental breakthrough computations of certain forms of scalar products \cite{1990_Slavnov_TMP_79_82} and of matrix elements of physical operators \cite{Maillet} from the underlying algebraic structure of the models (most economically expressed using the Algebraic Bethe Ansatz \cite{1979_Sklyanin, KorepinBOOK}).

A significant open problem, which has up to now received little attention, consists in asking how a generic but well-defined quantum state overlaps with eigenfunctions of a certain integrable model. For example, one might ask how a spin chain state built from well-defined spin projections at each site projects onto eigenstates of a Bethe Ansatz-solvable \cite{1931_Bethe_ZP_71} Hamiltonian. Such overlaps form the basic building blocks of a recently-proposed approach to address out-of-equilibrium dynamics in integrable systems \cite{2013_Caux_PRL_110}, but their calculation including the required knowledge of their scaling in the thermodynamic limit poses significant challenges. For the Lieb-Liniger Bose gas \cite{1963_Lieb_PR_130_1}, an analytic expression for certain overlaps (namely of the condensate-like state with eigenstates at finite interaction) has been recently found \cite{LLpaper}. Similar results for the XXZ model \cite{1958_Orbach_PR_112} are not yet known.

In this work we derive a practical determinant expression for the overlap of the N{\'e}el state with parity-invariant Bethe states of the XXZ model, starting from the original determinant formula obtained in Refs.~\cite{1998_Tsuchiya_JMathPhys_39,2012_Kozlowski_JSTAT_P05021, Pozsgay_1309.4593}. The advantage of our expression is that it can be numerically evaluated for large system sizes, and that its thermodynamic limit can be extracted analytically. Remarkably, it shows peculiar similarities with the determinant expression for the overlaps in the Lieb-Liniger model \cite{LLpaper}. Although we do not fully understand the origin of these similarities at this stage, this coincidence is most probably not fortuitous, and points to possible deep-rooted and `universal' links between quench situations in different models.

The paper is organized as follows. In Sec.~\ref{sec:XXZ} we review the standard Bethe Ansatz solution of the XXZ chain and we present the most important formulas related to the Bethe Ansatz that are needed in later chapters. In Sec.~\ref{sec:det_exp} we first present the final result and then we show in its subsection~\ref{sec:proof} the proof of the formula. We also discuss the limit to the isotropic spin chain in subsection~\ref{sec:XXX_limit}.

\section{The XXZ spin-1/2 chain}\label{sec:XXZ}
The one-dimensional spin-1/2 XXZ model is given by the Hamiltonian
\begin{equation}\label{eq:Hamiltonian_XXZ}
	H = J\sum_{j=1}^{N}\left(\sigma_{j}^{x}\sigma_{j+1}^{x}+\sigma_{j}^{y}\sigma_{j+1}^{y}+\Delta ( \sigma_{j}^{z}\sigma_{j+1}^{z}-1)\right)\epp
\end{equation}
The coupling constant $J>0$ fixes the energy scale and the parameter $\Delta$ describes the anisotropy of the nearest neighbour spin-spin coupling. The length of the chain is given by $N$ (which we choose to be even) and we impose periodic boundary conditions $\sigma_{N+1}^\alpha=\sigma_1^\alpha$, $\alpha=x,y,z$.
 
This Hamiltonian can be diagonalized by Bethe Ansatz \cite{1958_Orbach_PR_112}. We choose the ferromagnetic state $\left|\uparrow \uparrow \ldots \uparrow\right\rangle$ with all spins up as a reference state and construct interacting spin waves above this state. A state with $M$ down spins reads
\begin{subequations}\label{eq:BA_state}
\begin{equation}
	|\{\lambda_j\}_{j=1}^M\rangle = \sum\nolimits_{\{s_j\}_{j=1}^M \subset \{1,\ldots,N\}} \Psi_{M}\!\left(\{s_j\}_{j=1}^M|\{\lambda_j\}_{j=1}^M\right)\ \sigma_{s_1}^-\ldots\sigma_{s_M}^-|\uparrow\uparrow\ldots\uparrow\rangle
\end{equation}
with the explicit wave function in coordinate space 
\begin{equation} 
\Psi_{M}\!\left(\{s_j\}_{j=1}^M|\{\lambda_j\}_{j=1}^M\right)=\sum_{Q\in\mathcal{S}_M} (-1)^{[Q]} 
 \exp\left\{- i \sum_{a=1}^M s_a P(\lambda_{Q_a}) - \frac{i}{2} \sum_{\substack{ a,b=1\\ b>a}}^M \theta(\lambda_{Q_b} - \lambda_{Q_a}) \right\}\epp
\end{equation}
\end{subequations}
Here, the coordinates $s_j$, $j=1,\ldots,M$, denote the positions of the down spins, and we assume $s_j<s_k$ for $j<k$. The set $\mathcal{S}_M$ is the set of all permutations of integers $1, \ldots, M$, and $(-1)^{[Q]}$ denotes the parity of the permutation $Q$. Further, $P(\lambda)$ is the momentum associated to the rapidity $\lambda$,
\begin{subequations}
\begin{equation}
P(\lambda) = - i \ln \left[\frac{\sin(\lambda + i \eta/2)}{\sin(\lambda- i \eta/2)}\right]\epc
\end{equation}
and $\theta(\lambda)$ is the scattering phase shift given by
\begin{equation}
	\theta(\lambda) = i\ln\left(\frac{\sin(\lambda+i\eta)}{\sin(\lambda-i\eta)}\right)\epp
\end{equation}
\end{subequations}
The parameter $\eta$ is determined by the anisotropy parameter $\Delta=\cosh(\eta)$, and the set of rapidities $\{\lambda_j\}_{j=1}^M$  in Eqs.~\eqref{eq:BA_state} specifies the state. The latter is an eigenstate of the Hamiltonian \eqref{eq:Hamiltonian_XXZ}, and it is called Bethe state if the rapidities $\lambda_j$, $j=1,\ldots,M$, satisfy the Bethe equations
\begin{equation}\label{eq:BAE}
	\left(\frac{\sin(\lambda_j+i\eta/2)}{\sin(\lambda_j-i\eta/2)}\right)^N=-\prod_{k=1}^M\frac{\sin(\lambda_j-\lambda_k+i\eta)}{\sin(\lambda_j-\lambda_k-i\eta)}\:, \qquad j=1,\ldots,M \epp
\end{equation}
In this case the rapidities $\lambda_j$, $j=1,\ldots,M$, are roots of the function  
\begin{equation}\label{eq:func_a}
	\mathfrak{A}(\lambda) = 1+\mathfrak{a}(\lambda)
	\quad\text{with}\quad 
	\mathfrak{a}(\lambda)=\left(\frac{\sin(\lambda+i\eta/2)}{\sin(\lambda-i\eta/2)}\right)^N\prod\limits_{k=1}^M\frac{\sin(\lambda-\lambda_k-i\eta)}{\sin(\lambda-\lambda_k+i\eta)}\epc
\end{equation}
and they are called Bethe roots. The norm of a Bethe state is given by \cite{1981_Gaudin_PRD_23} 
\begin{subequations}\label{eq:norm_Bethe_state}
\begin{align}
\|\{\lambda_j\}_{j=1}^M\| &= \sqrt{\langle \{\lambda_j\}_{j=1}^M|  \{\lambda_j\}_{j=1}^M \rangle}\epc \\ 
\label{eq:norm_Bethe_state_b}
	\langle \{\lambda_j\}_{j=1}^M|  \{\lambda_j\}_{j=1}^M \rangle &= \sinh^M(\eta) \prod_{\substack{j,k=1\\j\neq k}}^M \frac{\sin(\lambda_j - \lambda_k + i \eta)}{\sin(\lambda_j - \lambda_k)} \det{}_{\!M} (G_{jk}) \epc\\
	G_{jk} &= \label{eq:Gaudin_matrix}  \delta_{jk}\left(NK_{\eta/2}(\lambda_j)-\sum_{l=1}^{M}K_\eta(\lambda_j-\lambda_l)\right) + K_\eta(\lambda_j-\lambda_k)\epc
\end{align}
\end{subequations}
where $K_\eta(\lambda)=\frac{\sinh(2\eta)}{\sin(\lambda+i\eta)\sin(\lambda-i\eta)}$ is the derivative of the scattering phase shift $\theta(\lambda)$.

In the following we will call states of the form \eqref{eq:BA_state} Bethe states or ``on-shell'' if the parameters $\{\lambda_j\}_{j=1}^M$ fulfill the Bethe equations \eqref{eq:BAE}. If they are arbitrary the state is called ``off-shell''. We call a state parity invariant, $|\{\pm\lambda_j\}_{j=1}^{M/2}\rangle$, if the rapidities fulfill the symmetry $\{\lambda_j\}_{j=1}^M= \{-\lambda_j\}_{j=1}^M = \{\lambda_j\}_{j=1}^{M/2}\cup \{-\lambda_j\}_{j=1}^{M/2} \equiv \{\pm\lambda_j\}_{j=1}^{M/2}$.

\section{Determinant expression for the overlaps with the N{\'e}el state}\label{sec:det_exp}
We are interested in the overlap of the zero-momentum N{\'e}el state given by
\begin{equation}\label{eq:Neel_state}
	|\Psi_0\rangle = \frac{1}{\sqrt{2}}\big(|\!\uparrow\downarrow \uparrow\downarrow\ldots\rangle+|\!\downarrow\uparrow\downarrow\uparrow \ldots\rangle\big)
\end{equation} 
with the XXZ Bethe states of the form \eqref{eq:BA_state}. Therefore we only consider Bethe states with $M=N/2$ flipped spins since the N{\'e}el state lies in this sector of the XXZ chain. For the sake of simplicity we choose $N$ divisible by four such that $M$ is even. We consider parity-invariant Bethe states $|\{\pm\lambda_j\}_{j=1}^{M/2}\rangle$ which have non-vanishing overlap $\langle \Psi_0 | \{\pm\lambda_j\}_{j=1}^{M/2}\rangle$. For the total momentum of these states we find 
\begin{subequations}
\begin{equation}
- i\sum_{j=1}^N \ln \left[\frac{\sin(\lambda_j + i \eta/2)}{\sin(\lambda_j - i \eta/2)}\right] = 0 \epc
\end{equation}
and all other odd conserved charges $\hat{Q}_{2n+1}$ \cite{Fagotti_2013} evaluate to zero as well,
\begin{equation}
\hat{Q}_{2n+1} |\{\pm\lambda_j\}_{j=1}^{M/2}\rangle = \sum_{j=1}^M P_{2n+1}(\lambda_j)  |\{\pm\lambda_j\}_{j=1}^{M/2}\rangle=0\epc
\end{equation}
\end{subequations}
because $P_{2n+1}$ is an odd function, $P_{2n+1}(\lambda) = i  \frac{\partial^{2n}}{\partial \mu^{2n}} \ln\left[ \frac{\sin(\lambda - \mu + i \eta/2)}{\sin(\lambda - \mu - i \eta/2) }\right]_{\mu\to 0}$.
That we are only interested in the overlap with parity-invariant Bethe states is motivated by the fact that the odd conserved charges, evaluated on non-parity-invariant Bethe states, are in general non-zero whereas their expectation value on the N{\'e}el state vanishes \cite{Fagotti_2013}. Most-recently it was proven that the overlap of the N\'eel state with non-parity-invariant Bethe states is indeed zero \cite{Brockmann_1403}.

For clarity, let us here quote the main result of our paper (whose derivation will be presented below): the normalized overlap of the zero-momentum N{\'e}el state $|\Psi_0\rangle$ with a parity-invariant Bethe state $|\{\pm\lambda_j\}_{j=1}^{M/2}\rangle$, which reads as follows:
\begin{subequations}\label{eq:overlap}
\begin{equation}
	\frac{\langle \Psi_0 | \{\pm\lambda_j\}_{j=1}^{M/2} \rangle }{\|\{\pm\lambda_j\}_{j=1}^{M/2}\|}=\sqrt{2} \left[\prod_{j=1}^{M/2}\frac{\sqrt{\tan(\lambda_j+i\eta/2) \tan(\lambda_j-i\eta/2)}}{2\sin(2\lambda_j)}\right]\sqrt{ \frac{\det_{M/2}(G_{jk}^{+})}{\det_{M/2}(G_{jk}^{-})}}
\end{equation}
where 
\begin{equation}\label{eq:overlap_b}
G_{jk}^\pm = \delta_{jk}\left(NK_{\eta/2}(\lambda_j)-\sum_{l=1}^{M/2}K_\eta^+(\lambda_j,\lambda_l)\right) + K_\eta^\pm(\lambda_j,\lambda_k)\epc \quad j,k=1,\ldots,M/2\epc
\end{equation}
\end{subequations}
$K_\eta^\pm(\lambda,\mu)=K_\eta(\lambda-\mu) \pm K_\eta(\lambda+\mu)$, and $K_\eta(\lambda)=\frac{\sinh(2\eta)}{\sin(\lambda+i\eta)\sin(\lambda-i\eta)}$. Note that here Bethe roots can be complex numbers (string solutions). The XXZ overlap formula \eqref{eq:overlap} for the N{\'e}el state looks very similar to the Lieb-Liniger overlap formula for a state which describes a Bose-Einstein condensate of one-dimensional free Bosons \cite{LLpaper}.

\subsection{Proof of an off-shell formula}\label{sec:proof}
In order to prove overlap formula \eqref{eq:overlap} we start with an expression for the overlap of the N{\'e}el state with an unnormalized off-shell state $|\{\tilde\lambda_j\}_{j=1}^{M}\rangle$ of the form \eqref{eq:BA_state}, which was proven in Ref.~\cite{2012_Kozlowski_JSTAT_P05021, Pozsgay_1309.4593} (see Eqs.~(2.26) -- (2.27) in Ref.~\cite{2012_Kozlowski_JSTAT_P05021} with $\xi=:-i \eta/2$). Introducing the short-hand notation $\sine{\alpha}{\beta} =\sin(\alpha+i\beta)$ these equations from Ref.~\cite{2012_Kozlowski_JSTAT_P05021} can be written as (note that $N=2M$)
\begin{subequations}\label{eq:overlap_S3}
\begin{align}\label{eq:overlap_S3a}
	\langle \Psi_0|\{\tilde{\lambda}_j\}_{j=1}^M\rangle &= \sqrt{2} \left[\prod_{j=1}^M\frac{\sine{\tilde\lambda_j}{+\eta/2}}{\sine{2\tilde\lambda_j}{0}}\;\frac{\sine{\tilde\lambda_j}{-\eta/2}^{M}}{ \sine{\tilde\lambda_j}{+\eta/2}^{M}} \right] \left[\prod_{j>k=1}^{M}\frac{\sine{\tilde\lambda_j+\tilde\lambda_k}{\eta}}{\sine{\tilde\lambda_j+\tilde\lambda_k}{0}}\right]\det{}_{\!M}(\delta_{jk}+U_{jk})\epc\\
	\label{eq:overlap_S3b}
	U_{jk} &= \frac{\sine{2\tilde\lambda_k}{\eta}\sine{2\tilde\lambda_k}{0}}{\sine{\tilde\lambda_j+\tilde\lambda_k}{0}\sine{\tilde\lambda_j-\tilde\lambda_k}{\eta}}\left[\prod_{\substack{l=1\\ l\neq k}}^{M}\frac{\sine{\tilde\lambda_k+\tilde\lambda_l}{0}}{\sine{\tilde\lambda_k-\tilde\lambda_l}{0}}\right]\left[\prod_{l=1}^M\frac{\sine{\tilde\lambda_k-\tilde\lambda_l}{-\eta}}{\sine{\tilde\lambda_k+\tilde\lambda_l}{+\eta}}\right]\left(\frac{\sine{\tilde\lambda_k}{+\eta/2}}{\sine{\tilde\lambda_k}{-\eta/2}}\right)^{2M}\epp
\end{align}
\end{subequations}
This expression is unhandy to perform the thermodynamic limit as well as for parity-invariant states $|\{\pm\lambda_j\}_{j=1}^{M/2}\rangle$ due to zeroes of the determinant and singularities in the prefactor. To perform the limit to parity-invariant states (not necessarily Bethe states) we set $\tilde\lambda_j = \lambda_j + \epsilon_j$ for $j=1,\ldots,M/2$ and $\tilde\lambda_j = -\lambda_{j-M/2} + \epsilon_{j-M/2}$ for $j=M/2+1,\ldots,M$. Here, the parameters $\lambda_j$, $j=1,\ldots,M/2$, are arbitrary complex numbers. We shall see that the main ingredients to the derivation of formula \eqref{eq:overlap} are the limits $\epsilon_j\to 0$, $j=1,\ldots,M/2$, and the pseudo parity invariance of the set $\{\tilde\lambda_j\}_{j=1}^{M} = \{\lambda_j+\epsilon_j\}_{j=1}^{M/2}\cup \{-\lambda_j+\epsilon_j\}_{j=1}^{M/2}$.  We derive then an off-shell version of Eq.~\eqref{eq:overlap}, which has the same form up to corrections which are zero when the rapidities satisfy the Bethe equations \eqref{eq:BAE}.

If we multiply the prefactor in \eqref{eq:overlap_S3a} with $\alpha_\text{reg}=\prod_{j=1}^{M/2}\left(\frac{\sine{2\epsilon_j}{0}}{\sine{0}{\eta}}\right)$ and the determinant by its inverse $\alpha_\text{reg}^{-1}$ we get regular expressions with well-defined limits $\epsilon_j\to 0$, $j=1,\ldots,M/2$, as well as a well-defined XXX scaling limit $\lambda\to\eta\lambda$ and $\eta\to 0$ afterwards.  Assuming an appropriate order of rapidities the lowest order in $\{\epsilon_j\}_{j=1}^{M/2}$ of the regularized prefactor and of the determinant read
\begin{subequations}
\begin{align}\label{eq:gamma}
	\gamma &= \sqrt{2}\left[\prod_{j=1}^{M/2}\frac{\sine{\lambda_j}{+\eta/2}\sine{\lambda_j}{-\eta/2}}{\sine{2\lambda_j}{0}^2}\right] \left[\prod_{\substack{j>k=1\\ \ \sigma=\pm}}^{M/2}\frac{\sine{\lambda_j+\sigma\lambda_k}{+\eta}\sine{\lambda_j+\sigma\lambda_k}{-\eta}}{\sine{\lambda_j+\sigma\lambda_k}{0}^2}\right]\epc\\
	\det{}_\text{reg} &= \lim\nolimits_{\{\epsilon_j\to 0\}_{j=1}^{M/2}} \left\{\prod_{j=1}^{M/2}\frac{\sine{0}{\eta}}{\sine{2\epsilon_j}{0}}\det{}_{\!M}(\delta_{jk}+U_{jk})\right\}\epp \label{eq:Dreg}
\end{align}
\end{subequations}
The task is now to calculate the limits $\epsilon_j\to 0$, $j=1,\dots,M/2$, in \eqref{eq:Dreg}. For this purpose we reorder the rows and columns of the matrix $\mathds{1}+U$ under the determinant in such a way that the pair $(\lambda_{k}+\epsilon_k,-\lambda_k+\epsilon_k)$ belongs to the two rows and two columns with indices $2k-1$ and $2k$.

We consider the resulting $M\times M$ matrix as a $M/2\times M/2$ block matrix built of $2\times 2$ blocks. Collecting all terms up to first order in all $\epsilon_j$, $j=1,\ldots,M/2$, we obtain on the diagonal the $2\times 2$ blocks 
\begin{subequations}
\begin{align}\label{eq:diagonal_block}
	1+U_{2k-1,2k-1} &=  1+\delta_k\frac{\sine{2\lambda_k}{+\eta}}{\sine{2\lambda_k}{0}}\mathfrak{a}_k\epc &     U_{2k-1,2k} &= -\frac{\sine{2\lambda_k}{-\eta}}{\sine{2\lambda_k}{+\eta}}\mathfrak{a}_k^{-1} + \delta_k\mathfrak{b}_k^{-} \epc \notag\\
	    U_{2k,    2k-1} &= -\frac{\sine{2\lambda_k}{+\eta}}{\sine{2\lambda_k}{-\eta}}\mathfrak{a}_k + \delta_k\mathfrak{b}_k^{+} \epc & 1+U_{2k,    2k} &= 1+\delta_k\frac{\sine{2\lambda_k}{-\eta}}{\sine{2\lambda_k}{0}}\mathfrak{a}_k^{-1}\epc \\
\intertext{and for the off-diagonal blocks ($1\leq j, k\leq M/2$, $j\neq k$)}
\label{eq:offdiagonal_block}
	U_{2j-1,2k-1} &=  \delta_k\frac{\sine{2\lambda_k}{+\eta}\sine{0}{\eta}}{\sine{\lambda_j+\lambda_k}{0}\sine{\lambda_j-\lambda_k}{+\eta}}\mathfrak{a}_k\epc & U_{2j-1,2k} &= \delta_k\frac{\sine{2\lambda_k}{-\eta}\sine{0}{\eta}}{\sine{\lambda_k-\lambda_j}{0}\sine{\lambda_j+\lambda_k}{+\eta}}\mathfrak{a}_k^{-1} \epc \notag\\	
	U_{2j,    2k-1} &= \delta_k\frac{\sine{2\lambda_k}{+\eta}\sine{0}{\eta}}{\sine{\lambda_j-\lambda_k}{0}\sine{\lambda_j+\lambda_k}{-\eta}}\mathfrak{a}_k \epc & U_{2j,    2k} &= \delta_k\frac{\sine{2\lambda_k}{-\eta}\sine{0}{\eta}}{\sine{\lambda_j+\lambda_k}{0}\sine{\lambda_k-\lambda_j}{+\eta}}\mathfrak{a}_k^{-1}\epc	
\end{align}
\end{subequations}
where we defined the abbreviations $\delta_k=\sine{2\epsilon_k}{0}/\sine{0}{\eta}$ for $k=1,\ldots,M/2$ and
\begin{equation}\label{eq:func_a_tilde}
	\mathfrak{a}_k = \tilde{\mathfrak{a}}(\lambda_k)= \left[\prod_{\substack{l=1\\ \ \sigma=\pm}}^{M/2}\frac{\sine{\lambda_k-\sigma\lambda_l}{-\eta}}{\sine{\lambda_k-\sigma\lambda_l}{+\eta}}\right]\left(\frac{\sine{\lambda_k}{+\eta/2}}{\sine{\lambda_k}{-\eta/2}}\right)^{2M}\epp
\end{equation}
Note that the function $\tilde{\mathfrak{a}}$ is different from the function $\mathfrak{a}$ of Eq.~\eqref{eq:func_a} because in the definition of $\mathfrak{a}$ the parameters $\{\lambda_k\}_{k=1}^{M}$ are Bethe roots whereas here in Eq.~\eqref{eq:func_a_tilde} they are arbitrary. The symbols $\mathfrak{b}_k^+$ and $\mathfrak{b}_k^-$ in Eq.~\eqref{eq:diagonal_block} denote the first order corrections of the elements $U_{2k,2k-1}$ and $U_{2k-1,2k}$, respectively. After a short calculation we obtain up to zeroth order in $\delta_k$
\begin{equation}
	\frac{\sine{2\lambda_k}{-\eta}}{\sine{2\lambda_k}{+\eta}}\mathfrak{a}_k^{-1}\mathfrak{b}_k^{+} + \frac{\sine{2\lambda_k}{+\eta}}{\sine{2\lambda_k}{-\eta}}\mathfrak{a}_k\mathfrak{b}_k^{-} = 2\cosh(\eta)-\sine{0}{\eta}\partial_{\lambda_k}\ln\Big\{\frac{\sine{\lambda_k}{+\eta/2}^{2M}}{\sine{\lambda_k}{-\eta/2}^{2M}}\prod_{\substack{l=1 \\ l\neq j}}^{M/2}\prod_{\sigma=\pm}\frac{\sine{\lambda_k+\sigma\lambda_l}{-\eta}}{\sine{\lambda_k+\sigma\lambda_l}{+\eta}}\Big\}\epp
\end{equation}
We further define $\alpha_k = \sqrt{-\frac{\sine{2\lambda_k}{+\eta}}{\sine{2\lambda_k}{-\eta}}\mathfrak{a}_k}$ and multiply the $M\times M$ matrix $\mathds{1}+U$ from the left and from the right respectively with the matrices
\begin{equation}
	\text{diag}_{M}\left(\alpha_1,\alpha_1^{-1},\ldots,\alpha_M,\alpha_M^{-1}\right)\epc \quad \text{diag}_{M}\left(\alpha_1^{-1},\alpha_1,\ldots,\alpha_M^{-1},\alpha_M\right)\epp 
\end{equation}
Since the determinants of these diagonal matrices are equal to one this transformation does not change the value of the determinant $\det_M(\mathds{1}+U)$. We see that the structure of the matrix becomes 
\begin{equation}\label{eq:matrix_structure}
	\left(\begin{array}{c@{\hspace{-3ex}}c@{\hspace{1ex}}c} 
	\left[\begin{array}{ll} 1-\delta_1\frac{\sine{2\lambda_1}{-\eta}}{\sine{2\lambda_1}{0}}\alpha_1^2 & 1+\delta_1\mathfrak{b}_1^{-}\alpha_1^2\\ 1+\delta_1\mathfrak{b}_1^{+}\alpha_1^{-2} & 1-\delta_1\frac{\sine{2\lambda_1}{+\eta}}{\sine{2\lambda_1}{0}}\alpha_1^{-2} \end{array}\right]&
	\delta_2\left[\begin{array}{cc}a_{12}&b_{12}\\ c_{12}&d_{12}\end{array}\right] & \dots \\[4ex]
	\delta_1\left[\begin{array}{cc}a_{21}&b_{21}\\ c_{21}&d_{21}\end{array}\right] &
	\left[\begin{array}{ll} 1-\delta_2\frac{\sine{2\lambda_2}{-\eta}}{\sine{2\lambda_2}{0}}\alpha_2^2 & 1+\delta_2\mathfrak{b}_2^{-}\alpha_2^2\\ 1+\delta_2\mathfrak{b}_2^{+}\alpha_2^{-2} & 1-\delta_2\frac{\sine{2\lambda_2}{+\eta}}{\sine{2\lambda_2}{0}}\alpha_2^{-2} \end{array}\right] &  \\
	\vdots&
	& \ddots \\
	\end{array}\right)\epc
\end{equation}
where the elements $a_{jk}$, $b_{jk}$, $c_{jk}$, and $d_{jk}$, $j,k=1,\ldots M/2$, of the off-diagonal blocks can be calculated by multiplying the $2\times 2$ block \eqref{eq:offdiagonal_block} with $\text{diag}_2(\alpha_j,\alpha_j^{-1})$ from the left and with $\text{diag}_2(\alpha_k^{-1},\alpha_k)$ from the right. 

Under the determinant the matrix \eqref{eq:matrix_structure} can be further simplified by replacing column $2k-1$ by the difference of columns $2k-1$ and $2k$ for all $k=1,\ldots, M/2$ and afterwards by replacing row $2j-1$ by the difference of rows $2j-1$ and $2j$ for all $j=1,\ldots, M/2$. Up to first order in each $\delta_1,\delta_2,\ldots, \delta_{M/2}$ the determinant of $\mathds{1}+U$ becomes 
\begin{equation}\label{eq:matrix_structure2}
	\det{}_{\!M}\left(\begin{array}{c@{\hspace{0ex}}c@{\hspace{0ex}}c@{\hspace{0ex}}c} 
	\left[\begin{array}{c@{\hspace{0.9ex}}c} \delta_1 D_{1}  & 0 \\ 0 & 1 \end{array}\right] &
	\left[\begin{array}{c@{\hspace{0.9ex}}c} \delta_2 e_{12} & 0 \\ 0 & 0 \end{array}\right] &
	\left[\begin{array}{c@{\hspace{0.9ex}}c} \delta_3 e_{13} & 0 \\ 0 & 0 \end{array}\right] & \dots \\[3ex]
	\left[\begin{array}{c@{\hspace{0.9ex}}c} \delta_1 e_{21} & 0 \\ 0 & 0 \end{array}\right] &
	\left[\begin{array}{c@{\hspace{0.9ex}}c} \delta_2 D_{2}  & 0 \\ 0 & 1 \end{array}\right] & 
	\left[\begin{array}{c@{\hspace{0.9ex}}c} \delta_3 e_{23} & 0 \\ 0 & 0 \end{array}\right] & \\[3ex]
	\left[\begin{array}{c@{\hspace{0.9ex}}c} \delta_1 e_{31} & 0 \\ 0 & 0 \end{array}\right] &
	\left[\begin{array}{c@{\hspace{0.9ex}}c} \delta_2 e_{32} & 0 \\ 0 & 0 \end{array}\right] & 
	\left[\begin{array}{c@{\hspace{0.9ex}}c} \delta_3 D_{3}  & 0 \\ 0 & 1 \end{array}\right] &	
	\\
	\vdots & & & \ddots
	\end{array}\right) =  \left[\prod_{j=k}^{M/2}\delta_k\right]\det{}_{\!M/2}
	\hspace{-0.5ex}\left[\begin{array}{c@{\hspace{1ex}}c@{\hspace{1ex}}c@{\hspace{1ex}}c} 
	D_{1}  & e_{12} & e_{13} & \dots \\
	e_{21} & D_{2}  & e_{23} & \\
	e_{31} & e_{32} & D_{3}  & \\[-0.7ex]
	\vdots & & & \ddots
	\end{array}\right]
\end{equation}
where $e_{jk}= a_{jk}-b_{jk}-c_{jk}+d_{jk}$. The diagonal elements $D_k$, $k=1,\ldots,M/2$, are given by 
\begin{align}
D_k &= 
	 \lim{}_{\{\delta_k\to 0\}_{k=1}^{M/2}}\left(-\frac{\sine{2\lambda_k}{-\eta}}{\sine{2\lambda_k}{0}}\alpha_k^{2} - \frac{\sine{2\lambda_k}{+\eta}}{\sine{2\lambda_k}{0}}\alpha_k^{-2} - \mathfrak{b}_k^{+}\alpha_k^{-2} - \mathfrak{b}_k^{-}\alpha_k^2 \right)\notag\\
& = \frac{\sine{2\lambda_k}{+\eta}}{\sine{2\lambda_k}{0}}\mathfrak{a}_k + \frac{\sine{2\lambda_k}{-\eta}}{\sine{2\lambda_k}{0}}\mathfrak{a}_k^{-1}+2\cosh(\eta)-\sine{0}{\eta}\partial_{\lambda_k}\ln\Big\{\frac{\sine{\lambda_k}{+\eta/2}^{2M}}{\sine{\lambda_k}{-\eta/2}^{2M}}\prod_{\substack{l=1 \\ l\neq k}}^{M/2}\prod_{\sigma=\pm}\frac{\sine{\lambda_k+\sigma\lambda_l}{-\eta}}{\sine{\lambda_k+\sigma\lambda_l}{+\eta}}\Big\} \notag\\
	&=
	\frac{\sine{2\lambda_k}{+\eta}}{\sine{2\lambda_k}{0}}\mathfrak{A}_k + \frac{\sine{2\lambda_k}{-\eta}}{\sine{2\lambda_k}{0}}\bar{\mathfrak{A}}_k +2M\sine{0}{\eta}K_{\eta/2}(\lambda_k)-\sum_{\substack{l=1\\ l\neq k}}^{M/2} \sine{0}{\eta}K_\eta^{+}(\lambda_k,\lambda_l)\epc
\end{align}
where we defined $K_\eta^{+}(\lambda,\mu)=K_\eta(\lambda-\mu)+K_\eta(\lambda+\mu)$, $K_\eta(\lambda)=\frac{\sine{0}{2\eta}}{\sine{\lambda}{+\eta}\sine{\lambda}{-\eta}}$ and $\mathfrak{A}_k=1+\mathfrak{a}_k$, $\bar{\mathfrak{A}}_k=1+\mathfrak{a}_k^{-1}$. Using these definitions the off-diagonal elements can be simplified to 
\begin{equation}
	e_{jk} = \sqrt{\frac{\sine{2\lambda_k}{+\eta}\sine{2\lambda_k}{-\eta}\mathfrak{a}_j}{\sine{2\lambda_j}{+\eta}\sine{2\lambda_j}{-\eta}\mathfrak{a}_k}}\left(K_{\eta}^{+}(\lambda_j,\lambda_k)+ f_{jk}\right)\epc
\end{equation}
where the symbols $f_{jk}$ are specified below (see Eq.~\eqref{eq:f_jk}). We can forget about the square roots as they cancel each other under the determinant. The factor $\prod_{k=1}^{M/2}\delta_k$ cancels exactly the factor $\alpha_\text{reg}^{-1}=\prod_{j=1}^{M/2}\frac{\sine{2\epsilon_j}{0}}{\sine{0}{\eta}}=\prod_{k=1}^{M/2}\delta_k^{-1}$ in Eq.~\eqref{eq:Dreg}. 

Hence, we reduced the $M$ dimensional determinant of overlaps with deviated parity-invariant states in the limit of vanishing deviations to an $M/2$ dimensional determinant which depends on $M/2$ independent parameters. The result can be eventually summed up as follows:
\begin{subequations}\label{eq:overlap_XXZ_offshell}
\begin{align}\label{eq:overlap_XXZ_offshell_a}
	 \langle \Psi_0 |\{\pm\lambda_j\}_{j=1}^{M/2}\rangle &= \left.\langle \Psi_0 |\{\lambda_j+\epsilon_j\}_{j=1}^{M/2}\cup \{-\lambda_j+\epsilon_j\}_{j=1}^{M/2}\rangle\right|_{\{\epsilon_j\to 0\}_{j=1}^{M/2}} \notag\\[1ex]
	  &=\left[\gamma\det{}_{\!M/2}(G_{jk}^{+}) + \mathcal{O}\left(\{\epsilon_j\}_{j=1}^{M/2}\right)\right]_{\{\epsilon_j\to 0\}_{j=1}^{M/2}} = \gamma\det{}_{\!M/2}(G_{jk}^{+})\epc
\end{align}
where the prefactor $\gamma$ is given by Eq.~\eqref{eq:gamma} and the matrix $G_{jk}^+$ reads
\begin{align}
G_{jk}^{+} &= \delta_{jk}\left(N\sine{0}{\eta}K_{\eta/2}(\lambda_j)-\sum_{l=1}^{M/2}\sine{0}{\eta}K_\eta^{+}(\lambda_j,\lambda_l)\right) + \sine{0}{\eta}K_\eta^{+}(\lambda_j,\lambda_k)\notag\\
&\qquad\quad + \delta_{jk}\frac{\sine{2\lambda_j}{+\eta}\,\mathfrak{A}_j+\sine{2\lambda_j}{-\eta}\,\bar{\mathfrak{A}}_j}{\sine{2\lambda_j}{0}} + (1-\delta_{jk})f_{jk}\:, \quad\qquad j,k=1,\ldots,M/2\\[3ex] \label{eq:f_jk}
f_{jk} 
&= \mathfrak{A}_k\left( \frac{\sine{2\lambda_j}{+\eta} \sine{0}{\eta}}{\sine{\lambda_j+\lambda_k}{0}\sine{\lambda_j-\lambda_k}{+\eta}} - \frac{\sine{2\lambda_j}{-\eta}\sine{0}{\eta}}{\sine{\lambda_j-\lambda_k}{0}\sine{\lambda_j+\lambda_k}{-\eta}} \right) + \mathfrak{A}_k\bar{\mathfrak{A}}_j\frac{\sine{2\lambda_j}{-\eta}\sine{0}{\eta}}{\sine{\lambda_j-\lambda_k}{0}\sine{\lambda_j+\lambda_k}{-\eta}} \notag\\
&\quad - \bar{\mathfrak{A}}_j\left(\frac{\sine{2\lambda_j}{-\eta}\sine{0}{\eta}}{\sine{\lambda_j-\lambda_k}{0}\sine{\lambda_j+\lambda_k}{-\eta}} + \frac{\sine{2\lambda_j}{-\eta}\sine{0}{\eta}}{\sine{\lambda_j+\lambda_k}{0}\sine{\lambda_j-\lambda_k}{-\eta}}\right)\epp
\end{align}
\end{subequations}
Note that here the matrix elements $G_{jk}^{+}$ contain an additional factor $\sine{0}{\eta}=i\sinh(\eta)$ compared to the earlier definition of $G_{jk}^{+}$ in Eq.~\eqref{eq:overlap_b}, which is convenient for taking the XXX limit $\eta\to 0$ which we shall do in Sec.~\ref{sec:XXX_limit}. Formula \eqref{eq:overlap_XXZ_offshell} holds for arbitrary complex numbers $\lambda_j$, $j=1,\ldots,M/2$.

If we are on-shell in Eq.~\eqref{eq:overlap_XXZ_offshell}, i.\,e.~if the set $\{\pm\lambda_j\}_{j=1}^{M/2}$ satisfies the Bethe equations \eqref{eq:func_a}, and if we are not in the XXX limit with rapidities at infinity, $\mathfrak{A}_j$ and $\bar{\mathfrak{A}}_j$ just vanish. Together with the norm of Bethe states and using the symmetry of the Gaudin matrix \eqref{eq:Gaudin_matrix} as well as the relation 
\begin{equation}
	\det{}_{\!M}\left(\begin{array}{cc}A & B\\ B & A\end{array}\right)= \det{}_{\!M/2}(A+B)\det{}_{\!M/2}(A-B)
\end{equation}
for block matrices we finally gain the overlap formula~\eqref{eq:overlap}, where the additional factors $\sine{0}{\eta}$ in $G_{jk}^{+}$, $1\leq j,k \leq M/2$, cancel against the prefactor $\sqrt{\sine{0}{\eta}^M}$ in the norm formula \eqref{eq:norm_Bethe_state}.

\subsection{The XXX limit including rapidities at infinity}\label{sec:XXX_limit}
At the isotropic point $\Delta=1$, there are Bethe states with rapidities at infinity that need to be specially treated. In order to obtain an appropriate overlap formula for the XXX case including such rapidities at infinity we use the off-shell formula \eqref{eq:overlap_XXZ_offshell}. As long as we are off-shell we can scale all rapidities by $\eta$, send $\eta$ to zero and can afterwards send some of the rapidities to infinity. Within this scaling Eq.~\eqref{eq:overlap_XXZ_offshell_a} reduces to 
\begin{subequations}
\begin{equation}\label{eq:overlap_XXX_offshell}
	\langle \Psi_0 |\{\pm\lambda_j\}_{j=1}^{M/2} \rangle = \tilde\gamma\det{}_{\!M/2}(\tilde{G}_{jk}^{+})\:,
\end{equation}
where now
\begin{align}
 \tilde\gamma &= \sqrt{2}\left[\prod_{j=1}^{M/2}\frac{\lambda_j^2+\frac{1}{4}}{4\lambda_j^2}\right]
\left[\prod_{j>k=1}^{M/2}\prod_{\sigma=\pm}\frac{(\lambda_j+\sigma\lambda_k)^2+1}{(\lambda_j+\sigma\lambda_k)^2}\right]\\ \label{eq:GQ_offshell}
\tilde{G}_{jk}^{+} &= \delta_{jk}\left(2M\tilde{K}_{1/2}(\lambda_j)-\sum_{l=1}^{M/2}\tilde{K}_1^{+}(\lambda_j,\lambda_l)\right) + \tilde{K}_1^{+}(\lambda_j,\lambda_k) \quad\qquad [j,k=1,\ldots,M/2]\notag\\
&\qquad\quad + \delta_{jk}\frac{(2\lambda_j+i)\,\mathfrak{A}_j+(2\lambda_j-i)\,\bar{\mathfrak{A}}_j}{2\lambda_j} + (1-\delta_{jk})\tilde{f}_{jk}
\epc
\end{align}
\end{subequations}
$\tilde{K}_\alpha^{+}(\lambda,\mu)=\tilde{K}_\alpha(\lambda-\mu)+\tilde{K}_\alpha(\lambda+\mu)$, and $\tilde{K}_\alpha(\lambda)=\frac{2\alpha}{\lambda^2+\alpha^2}$. The explicit form of $\tilde{f}_{jk}$ is not interesting since either the $\tilde{f}_{jk}$ vanish due to Bethe equations or they give subleading corrections if one of the parameters $\lambda_j$ goes to infinity. The term $\mathfrak{a}_j$ reads now
\begin{equation}
	\mathfrak{a}_j = \left[\prod_{k=1}^{M/2}\prod_{\sigma=\pm}\frac{\lambda_j-\sigma\lambda_k-i}{\lambda_j-\sigma\lambda_k+i}\right]\left(\frac{\lambda_j+i/2}{\lambda_j-i/2}\right)^{2M}\epc
\end{equation}
and we have again $\mathfrak{A}_j=1+\mathfrak{a}_j=0$ if the set $\{\pm\lambda_j\}_{j=1}^{M/2}$ satisfies the Bethe equations of the XXX model.

Let us consider the case when $n$ pairs of Bethe roots are at $\pm\infty$ in such a way that the difference and the sum of two Bethe roots which do not belong to the same pair is also infinity. Let us denote the corresponding parameters by $\mu_j$, $j=1,\ldots,n$. We have $m=M/2-n$ finite pairs $(\lambda_j,-\lambda_j)$ that fulfill the Bethe equations
\begin{equation}
	\mathfrak{a}_j = \left[\prod_{k=1}^{m}\prod_{\sigma=\pm}\frac{\lambda_j-\sigma\lambda_k-i}{\lambda_j-\sigma\lambda_k+i}\right]\left(\frac{\lambda_j+i/2}{\lambda_j-i/2}\right)^{2M} = -1\:, \qquad j=1,\ldots,m\epp
\end{equation}
We used the fact that all factors including one of the parameters $\mu_k$, $k=1,\ldots,n$, are equal to one. The Bethe equations for all infinite Bethe roots are trivial. The prefactor $\tilde\gamma$ has no divergencies in $\mu_j$ and just reads 
\begin{equation}
\tilde\gamma = \frac{\sqrt{2}}{4^n}\left[\prod_{j=1}^{m}\frac{\lambda_j^2+1/4}{4\lambda_j^2}\right]\left[\prod_{j>k=1}^{m}\prod_{\sigma=\pm}\frac{(\lambda_j+\sigma\lambda_k)^2+1}{(\lambda_j+\sigma\lambda_k)^2}\right].
\end{equation}
Now, we take the limits $\mu_j\to\infty$, $j=1,\ldots,n$, in Eq.~\eqref{eq:GQ_offshell} and treat all terms containing $\mathfrak{A}_j$ or $\bar{\mathfrak{A}}_j$ carefully. The determinant simplifies to 
\begin{equation}
\lim\nolimits_{\{\mu_j\to\infty\}_{j=1}^n}\left(\frac{\det{}_{\!M/2}(\tilde{G}_{jk}^{+})}{\mathcal{N}(\{\mu_j\}_{j=1}^{n})}\right) = 4^n(2n)!\det{}_{\!m}(\hat G_{jk}^{+})
\end{equation}
where $\hat{G}_{jk}^{+}$ is a reduced $m\times m$ version of the $M/2\times M/2$ matrix $\tilde{G}_{jk}^{+}$ in Eq.~\eqref{eq:GQ_offshell},
and where we used a factor $\mathcal{N}(\{\mu_j\}_{j=1}^{n})=\prod_{j=1}^n\mu_j^{-2}$ to get a non-vanishing result. 

We need the same factor $\mathcal{N}(\{\mu_j\}_{j=1}^{n})$ to correctly renormalize the norm formula~\eqref{eq:norm_Bethe_state}.  We introduce the abbreviations $\Lambda_{\pm}^{\!2m} =  \{\pm\lambda_j\}_{j=1}^{m}$,  $\mathcal{M}_{\pm}^{2n}=\{\pm\mu_j\}_{j=1}^{n}$, and $\Lambda_{\pm}^{M} = \Lambda_{\pm}^{\!2m} \cup \mathcal{M}_{\pm}^{2n}$ where the subscripts $\pm$ and $2m$, $2n$ are reminiscent to the parity invariance of the state and the $2m$ flipped spins, $2n$ pairs at infinity, respectively. The square of the norm of the state can be written as
\begin{equation}
	\frac{\langle \Lambda_{\pm}^{M}|\Lambda_{\pm}^{M}\rangle}{\mathcal{N}^2(\{\mu_j\}_{j=1}^{n})} \Big|_{\{\mu_j\to\infty\}_{j=1}^n} 
 = \langle \Lambda_{\pm}^{\!2m}|\left(S^+\right)^{2n}\left(S^-\right)^{2n}|\Lambda_{\pm}^{\!2m}\rangle = (4n)!\: \langle \Lambda_{\pm}^{\!2m} |\Lambda_{\pm}^{\!2m}\rangle\epp
\end{equation}
The norm square of the Bethe state $|\Lambda_{\pm}^{\!2m}\rangle$ is given by the XXX limit of the norm formula~\eqref{eq:norm_Bethe_state_b} and reads
\begin{equation}
	\langle \Lambda_{\pm}^{\!2m} |\Lambda_{\pm}^{\!2m}\rangle\ = \left[\prod_{j=1}^m \frac{\lambda^2_j+1/4}{\lambda^2_j}\right]  \left[\prod_{j>k=1}^m\prod_{\sigma=\pm}\frac{((\lambda_j+\sigma\lambda_k)^2+1)^2}{(\lambda_j+\sigma\lambda_k)^4}\right] \det{}_{\!2m}(\hat{G}_{jk})\epc
\end{equation}
where $\hat{G}_{jk}$ is a reduced $2m\times 2m$ version of the $M\times M$ Gaudin matrix $\tilde{G}_{jk}$. 

All together we obtain for the overlap of the N{\'e}el state \eqref{eq:Neel_state} with a normalized, parity-invariant XXX Bethe state $|\Lambda_{\pm}^{(m,n)}\rangle$, where $m$ pairs $(\lambda_j,-\lambda_j)$ are finite and all other rapidities $\pm\mu_j$, $j=1,\ldots,n$, are at $\pm\infty$ ($N_{\infty}=2n$, $2m+2n=M=N/2$):
\begin{subequations}\label{eq:overlap_XXX_onshell}
\begin{align}
	&\langle \Psi_0 |\Lambda_{\pm}^{(m,n)}\rangle = \left.\frac{\langle \Psi_0 |\Lambda_{\pm}^{M}\rangle }{ \|\Lambda_{\pm}^{M}\|}\right|_{\{\mu_j\to\infty\}_{j=1}^{n}} = \frac{\sqrt{2}\,N_{\infty}!}{\sqrt{(2N_{\infty})!}}
	\left[\prod_{j=1}^m \frac{\sqrt{\lambda^2_j+1/4}}{4\lambda_j}\right]  
	\sqrt{\frac{\det{}_{\!m}\hat{G}_{jk}^{+}}{\det{}_{\!m}\hat{G}_{jk}^{-}}}\:,\\
	&\hat{G}_{jk}^{\pm} = \delta_{jk}\left(2M\tilde{K}_{1/2}(\lambda_j)-\sum_{l=1}^{m}\tilde{K}_1^{+}(\lambda_j,\lambda_l)\right) + \tilde{K}_1^{\pm}(\lambda_j,\lambda_k)\:, \quad j,k=1,\ldots,m
\end{align}
\end{subequations}
with $\tilde{K}_\alpha^{\pm}(\lambda,\mu)=\tilde{K}_\alpha(\lambda-\mu)\pm \tilde{K}_\alpha(\lambda+\mu)$ and $\tilde{K}_\alpha(\lambda)=\frac{2\alpha}{\lambda^2+\alpha^2}$.
Note again that in Eqs.~\eqref{eq:overlap_XXX_onshell} Bethe roots can be complex numbers (string solutions).

\section{Summary}
In this work we obtained an exact expression for the overlap of the zero-momentum N{\'e}el state with parity-invariant Bethe states of the XXZ model for any value of the aniso\-tropy~$\Delta$. Generalizations to other simple classes of initial states (dimer and $q$-dimer states) as in Ref.~\cite{Pozsgay_1309.4593} are straightforward. Our result has a remarkable ``Gaudin-like'' form, which has also recently been found in the Lieb-Liniger case \cite{LLpaper}, and suggests a potentially deep link between such overlaps in integrable systems. On one hand, it allows for a rigorous proof of the Lieb-Liniger overlap formula \cite{Brockmann_2014}. On the other hand, the fact that our formula applies in the thermodynamic limit $N \to \infty$ opens the possibility of using the method of Ref.~\cite{2013_Caux_PRL_110} to investigate out-of-equilibrium dynamics in spin chains, paralleling the results of Ref.~\cite{LLpaper}. Other applications include the comparison with asymptotic results for the dynamical free energy \cite{Balasz_Dynamical}, and finally taking the infinite Trotter number limit for the surface free energy of an open spin chain as in Ref.~\cite{2012_Kozlowski_JSTAT_P05021}. We will investigate these topics in future publications.

\section*{Acknowledgements}
We acknowledge useful discussions with Davide Fioretto, Bal{\`a}zs Pozsgay, Pasquale Calabrese. 
We acknowledge support from the Foundation for Fundamental Research on Matter (FOM) and the Netherlands Organisation for Scientific Research (NWO).

\end{document}